# TO PARTICIPATE OR NOT TO PARTICIPATE: AN INVESTIGATION OF STRATEGIC PARTICIPATION IN STANDARDS

Paras Bhatt[1], Claire Vishik[2], Govind Hariharan[3], H. Raghav Rao[1]

**1** The University of Texas at San Antonio
**2** Intel Corporation
**3** Kennesaw State University

*Abstract*

*Essential functionality in the ICT (Information and Communication Technology) space draws from standards such as HTTP (IETF RFC 2616, Bluetooth (IEEE 802.15) and various telecommunication standards (4G, 5G). They have fuelled rapid growth of ICT sector in the last decades by ensuring interoperability and consistency in computing environment. Research shows that firms that backed ICT standards and participated in standards development, have emerged as industry innovators. Standards development thus clearly has benefits for participating companies as well as technology development and innovation in general. However, significant costs are also associated with development of standards and need to be better understood to support investment in standardization necessary for today's ICT environment. We present a conceptual model that considers the potential for market innovation advantage and influence on innovation and efficiency from standardization work, to build a forward-looking decision model that can guide an organization's standards development activities. We investigate and formalize motivations that drive firms to participate in standardization, specifically, influence of standards and market innovation advantages. Our model can serve as a strategic decision framework to drive assessments of a firm's participation in standards development. We test our model with a use case on an established access control approach that was standardized more than two decades ago, Role Based Access Control (RBAC) using historical data. The investigation of the case study shows that influence and market innovation advantage are significant indicators of success in standards development and are viable criteria to model a firm's decision to participate (or not to participate) in a specific area of standardization.*

## A. Introduction

Wi-fi (and connectivity in general) has become synonymous with the internet and is used to refer to internet usage. When people say "wi-fi" is working, they don't realize that they are referring to one of the fundamental connectivity standards that shaped the mobile Internet. The standardization of essential features of the ICT space supports interoperability makes large-scale production of equipment based on these standards possible (Wang et al, 2016). Using identical standards and technical specifications makes ICT products interoperable and reliable worldwide (Agard and Kusiak, 2004). Standards support emergence and deployment of innovation (Jiang et al, 2018a), enable emergence of newer markets and reignite demand in saturated markets (Tassey, 2002). According to Neshati and Daim's (2017) the major drivers for standards participation are economic, strategic, organizational and legal. Simcoe (2003) explore how firms' benefits from participation is related to the speed of the decision to participate. Tao et al., (2005) looked at managing intellectual assets for maximum value extraction from standards. Ancona and Caldwell (1992) show how innovation in organizations can be driven through interactions with standards and standards setting bodies. Baird (2007) also show that industry is best equipped to develop viable standards compared to governments. Through participation in standardization firms can reap economic benefits, extract maximum value through innovation, and build strategic partnerships through standards development.

Participation in standards is a substantial commitment for companies due to R&D costs, time-commitments, and uncertainty about adoption associated with standards' development (Nelson et al, 2005). These give rise to a dilemma: whether firms should participate in standards development or not. Since significant benefits, costs and uncertainties are associated with the decision to participate (Hill, 1997), it becomes important for companies to conduct a thorough analysis. Previous studies examine how product/patent-based strategies affect companies' motivation to participate in standardization (Baron et al., 2019). However, they do not cover influence of standards on market innovation advantages or market growth through innovation. The scope of standards development, in terms of value of a standard or associated emerging or mature technology (Smet et al., 2017) has seen limited research from economic point of view. Another area with limited research is how standards influence and affect changes in market innovation advantage during its lifecycle. To model how standards influence affects market innovation advantage during a standards lifecycle, we use modified Gompertz function (Kececioglu et al., 1994) and apply it to the decision to participate in standards development.

A criterion for deciding to pursue standards is market innovation advantage that it affords a participating company (Yeniyurt et al, 2005). This can be presented in terms of benefits acccruing to a company through standardization such as streamlined production (Farell and Saloner, 1985), interoperability of products, and ability to enter new markets with positive effects on innovation, quality, and growth. Another criterion is strategic influence on innovation that standards confers on the firm. This relates to increase in knowledge/expertise over time (Ryu et al, 2005) and growth of companies adopting a standard (O'Connor and Loomis, 2010). These criteria consider costs, market innovation advantage, and standards' lifecycle.

Our model determines the decision to participate in standards development by considering several criteria, like its market innovation advantage, influence on other firms, and probability of future events in standard's lifecycle. Using variables that define influence and market innovation advantage, we establish criteria and associated thresholds for standards participation. We pursue following research questions:

**RQ1** - What are the criteria considered by a company in their decision to participate (or not to participate) in standards' development?

**RQ2** – Given the above criteria, how can the decision to participate (or not to participate) in a standardization effort be modeled?

We examine a use-case study-based analysis of a historical American National Standards Institute (ANSI) standard – Role Based Access Control (RBAC) (Computer Security Resource Center, 2020), which was developed by National Institute of Standards and Technology (NIST) (O'Connor and Loomis, 2010). Our model illustrates market innovation advantage and influence (on innovation) are significant indicators of success that can drive the decision to participate. Companies gained benefits when they backed certain standards (RBAC) over competing standards or did not participate in standardization at all. The rest of the paper is as follows: Section B, highlights key literature in standards development. Section C and D, outlines our model and use case assessment. Finally, we discuss contributions from this research and highlight our future plans.

## B. Literature Review

As ICT advanced significantly over last several decades, standardization retained its importance. Other industries outside of the ICT sector engaged in ICT standardization as well (Folmer and Jakobs, 2020) as the impact of these standards expanded to other sectors. For example, ISO 20022 is used in SWIFT for enabling international financial transactions (Scott and Zachariadis, 2012). The United States Standards Strategy (USSS 2020) states that "Standards are the infrastructure for innovation, a critical component in bringing technologies from the lab to the market. Standards create a common language for trade and improve quality of life by enhancing safety, security, interoperability and the environment."[1]

Traditional standard setting may involve forming industry alliances to pursue shared development of standards creating a competition of ideas and approaches. Researchers note that these alliances may not always focus on the progress of an industry sector. (Leiponen, 2008). The proliferation of open standards and open-source architectures led to the emergence of approaches that are not limited to conventional standards-setting activities (Bonaccorsi et al., 2006). Also, the focus of participation in standardization varies between firms pursuing open standards development and those engaged in private standards development models. Though there have been studies that address standards development within a shared private-open ecosystems (W3C Encrypted Media Extensions - EME), some researchers caution against mixing the two models for various reasons (Halpin, 2017).

Libicki (2016) note that standards are analogous to language and enable efficient communication in the IT sector, alluding to easier design and off-the-shelf interoperability of ICT products. The development of standards helps firms achieve economies-of-scale and mass customization (Wang et al, 2016). Standards influence not only the micro economy of a company, but also have macro effects on a nation's economy. Grimes and Yang's (2018) work on global technology transfer notes how nation economies are built through strong ICT.

Firms can strategically build influence on innovation with help of their leadership in standardization. Attracting greater investment, adding suppliers and consumers, accruing and speeding market innovation advantages through market growth, quality, and efficiency are some examples of how standards and standardization affords influence on a company (Hebb and Wójcik, 2005). Standards also influence the suppliers and partners a company attracts (Soh 2009). The successful development of open standards could afford companies' greater efficiency in production, manufacturing, and trade. For example, a successful standard for video graphics performance with broad adoption is likely to impact manufacturing of graphic cards and GPUs. Successful forward-looking standards with broad adoption influence the industry as a whole (Wigand et al, 2005). Therefore, the influence on (broadly understood) innovation is frequently a key determinant of its success.

---

[1] Dr. Walter G. Copan, Under Secretary of Commerce for Standards and Technology and Director of the National Institute of Standards and Technology, 2020.

In addition to influence on the direction of innovation, standards result in various market and economic benefits (Marinagi et al, 2014). Eisend et al (2016) note that market innovation advantage and technology significantly influence new product performance, result in greater market share and ensure better service quality. In ICT sector, firms actively engaged in developing, deploying and maintaining standards have seen significant growth in market success innovation and performance. Market innovation advantage of firms with strong ICT is inherently linked to their knowledge competencies. Globally accepted ICT standards help companies achieve such advantage. Firms developing standards have greater impact on innovation than adopters of the standard (Yeniyurt et al, 2005). There is a reciprocal effect between technology development and technology standardization (Jiang et al, 2018b). As standards are developed and continually updated over its lifecycle, they foster greater innovation and subsequent advancements in use of underlying technology governing the standard.

The knowledge garnered from standards can be utilized to develop complementary products and ensure wider distribution of both the standards and its underlying technology. Ryu et al (2005) study how knowledge in enterprise information portals is acquired through learning processes. Investment in standards create specialized knowledge about the efficient use of technology and innovation thus enabling accelerated adoption of such standards on a global scale.

While ensuring high quality technology definition and interoperability for consumers across the world, standards also have significant implications for market evolution. The development of standards affect the market structure and guide decisions to enter new markets (Tassey, 2000). Standards help a company to reinvigorate saturated markets, explore niche markets and lead to creation of new markets. Soh (2009) notes that company network patterns are representative of market innovation advantage. This follows the approach in the paper which considers the positive effects of competition of standardization ideas. The presence of competing standards affects overall market innovation advantage accrued from a standard and tends to enhance the quality and potential of technical standards.

The success and failure of a standard is also affected by its lifecycle. Considerable initial investments may be required during initiation phase of standards development, with additional capital being required during different iterations. Since upfront costs are involved in standards development, the probability of future success or failure should be modelled when taking into consideration the decision to participate in standards development. Based on these metrics of standards, we put forth our model of the criteria that can drive the decision to participate in standards development.

## C. Methodology

We focus on three different sets of metrics important for standards development in a firm. First, costs associated with development like labor costs for technical personnel, R&D and membership fees required for joining a standards consortium. Second, benefits that standards afford organizations with regard to innovation, user adoption, and efficiency. The benefits can be studied from a dual perspective of influence (on innovation) and market innovation advantage. Lastly, since we examine forward-looking decisions, the risk that is entailed while making decisions about the future should also be considered. A benefit-cost-risk analysis is an effective framework for evaluating performance of a standard.

The influence a standard has on current and future technology development, and the market innovation advantages from developing the standard is important for its success. The ability to impact industry through innovation and enter new markets combine to form benefits that developing broadly adopted standards can afford a participating firm.

Time is another important factor and based on performance during different phases in standards development, the decision to increase, decrease or exit participation, and the decision related to associated funding can be made by firms. Finally, we incorporate risk in our model to enable an accurate estimation of future scenarios and the resulting decision to participate that can be modified at different stages in a standard's lifecycle.

In Table 1 and Figure 1, we explain our **I**nfluence and **M**arket **I**nnovation **A**dvantage for **P**articipation in **S**tandards (I-MIAPS) Model that includes criteria for standards development and broad set of variables for measuring its influence and charting its progress. We follow a market economics approach to build our model on two important standard criteria – influence and market innovation advantage. In terms of market innovation advantage, we further undertake a cost-benefit approach to estimating the current and future advantages. The model considers costs, influence and market innovation advantage with respect to time to obtain a decision to participate.

In a typical scenario:

- Costs (C) include initial investment and capital infusions (V), labor costs (L) and R&D costs (R); (Nelson et al., 2005; O'Connor and Loomis, 2010).
- At time (t) Market innovation (dis) advantage (M(t)) includes change in market coverage (ΔMC), change in partners and suppliers (ΔSB), change in compliant products (ΔCP); (Tassey, 2002; Yeniyurt et al, 2005; Soh, 2009; Marinagi et al., 2014; Eisend et al., 2016; Wang et al., 2016).
- Standards' Influence ($I_t$) is given by the number of companies that adopt a standard ($k_t$); (Hill, 1997; O'Connor and Loomis, 2010).
- At different stages of the standards lifecycle, a modified Gompertz function based on standards influence determines the maximum market innovation (dis) advantage to a firm from adopting the standard (Kececioglu et al., 1994).

These metrics can provide a snapshot of standards performance and drive the decision to participate.

| Model | **Decision to participate** = $f$ (Costs (C), Influence ($I_t$), Market Innovation Advantage ($M_t$)) |
|---|---|
| **Components** | **Variables** |
| **Cost (C)** | $C = f(V, L, R)$<br>Level of initial investment and sustained capital infusion (*V*), costs associated with labor, membership fees, technical personnel (*L*) or R&D (*R*) |
| **Market Innovation (dis) Advantage** | $\Delta M(t) = f(\Delta MC, \Delta SB, \Delta CP)$<br>Change in market coverage and innovation (Δ*MC*), new partnerships and change supplier base (Δ*SB*), increase in interoperable compliant (with the standard) products (Δ*CP*). |
| **Influence ($I_t$)** | $I_t = f(1/k_t)$<br>$k_t$ is the number of companies that adopt a standard at time *t* |
| **Standards Life cycle (t)** | $M = d + ab^{kt}$<br>During a standard's lifecycle, a modified Gompertz model (Kececioglu et al., 1994) based on market innovation advantage (M) and standards influence (k) determines the maximum advantage to a firm (*d + a*) as the standard reaches maturity T → ∞. The initial (dis) advantage when the firm first decides to participate in the standard at T = 0, is given by (*d + ab*). Small values of c indicate greater advantage as a small number of firms adopt and therefore share benefits from the knowledge acquired through standards. As more firms join, market innovation advantage is distributed across greater number of firms as the standard reaches maturity. |

**Table 1:** *I-MIAPS Model Components*

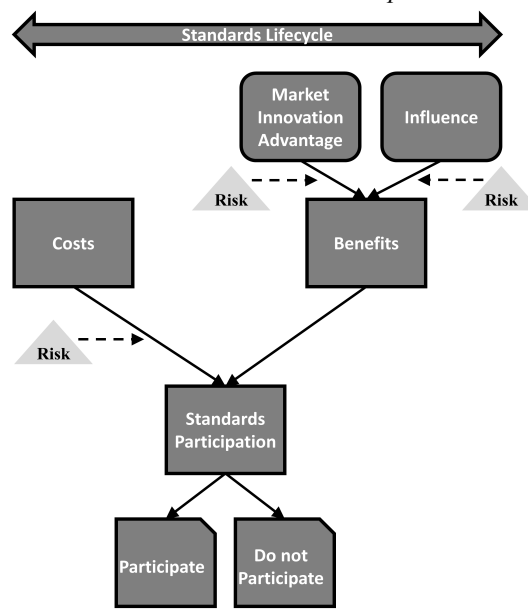

**Figure 1:** *I-MIAPS Model*

We incorporate risks (depicted by circles in Figure 2) pertaining to future costs, benefits and risk of failure of a standard in our decision model. Since from the inception of a standard and its first technical specification to its ultimate acceptance as a codified standard, there are several intermittent steps involved and thus it is necessary to consider how lifecycle of a standard affects decision to participate. For example, standards require upfront fixed investment whereas returns generally do not accrue outright. These returns can be obfuscated by complexities associated with development and adoption of technical standards. Successes in development and adoption of standards are revealed at various points in a standard's lifecycle. Thus, a company needs to continually assess their level of participation and make relevant decisions – no participation, full participation, and moderate/minimal participation. To capture standards' lifecycle aspects, we use a modified Gompertz function (Kececioglu et al., 1994) to incorporate time and influence to address the needs of a firm to monitor its involvement in standards and use our model to make relevant decisions - to invest, to increase investment, to continue, to exit and to reject participation. If a firm decides not to participate, it incurs an opportunity cost as there are likely to be other firms that participate and develop the standard. The terminal value in 'do not participate' considers such costs.

These decisions can be useful for analyzing both new and existing standard development activities of a firm. Figure 2 presents decision scenarios with possible outcomes in the I-MIAPS model.

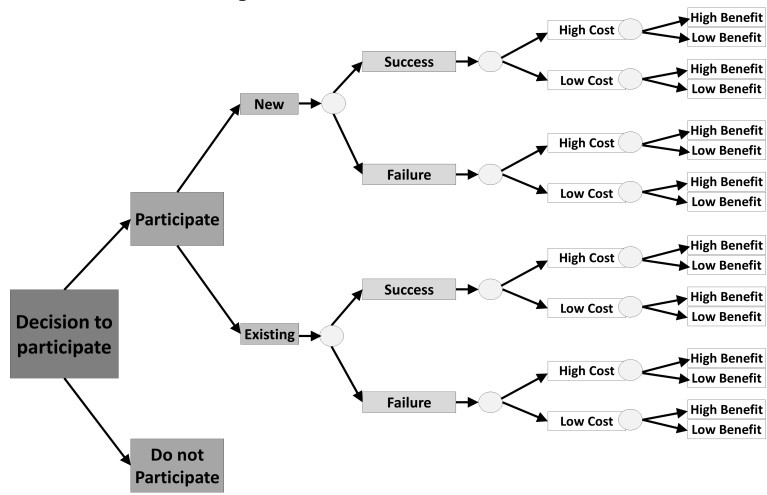

**Figure 2:** *The Basic Decision Model for Standards Participation*

We can have several scenarios regarding standards participation, see Figure 3. We include costs, benefits, influence, and risks associated with standards development and chart the favorable scenarios. There are clear choices for participation. However, level of participation can vary across participants as it is dependent on risk and investment strategy of individual firms. In this way, the I-MIAPS model incorporates strategic differences among companies that wish to use this model. The four quadrants derive empirical values from our decision model explained in Figure 2.

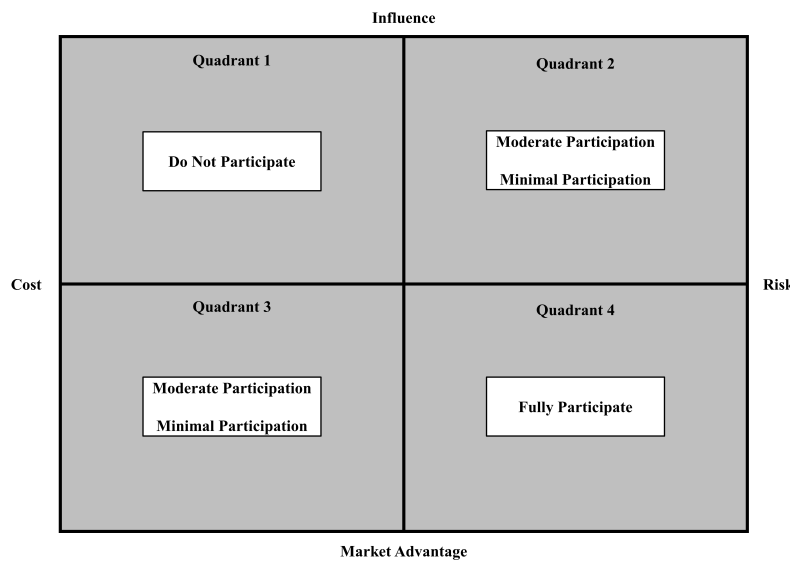

**Figure 3:** *Participation Scenario Matrix for Standards Development*

## D. Use Case Assessment

We conduct a use case assessment from O'Connor and Loomis (2010) evaluating performance of the RBAC standard. Using historical analysis, we build a forward-looking model that incorporates not only retrospective analysis but also a prospective one for future standards development. Studying a historical scenario provides important lessons, some public data, and benefit of seeing how standards evolve. We investigate the role of influence and market innovation advantage of RBAC standard over time. We apply our model to this use case and study standards participation. The I-MIAPS model differs from the use case by analyzing standards influence rather than net economic benefits. This differentiation is necessary because the I-MIAPS model not only considers historical standards development, but can also be used to guide new standards development.

RBAC was developed by NIST. During early 1990s, due to lack of a unified standard, individual firms could not develop interoperable access control products. The available options, Mandatory Access Control (MAC) and Discretionary Access Control (DAC), were not flexible to handle evolving business needs. In response, NIST developed RBAC model in 1992 which stimulated work on RBAC. Without active participation of NIST, RBAC would have developed at a much slower rate and with considerably higher costs.

RBAC standardization work resulting in a series of standards that continue to be updated, including INCITS 359-2012 (R2017), contributed to upward shift in value-added for companies that adopted these standards. NIST's involvement significantly improved efficiency of R&D, production, and commercialization of products and technologies based on RBAC. RBAC approach gained prominence in the 1990s replacing some uses of the Access Control Lists (ACLs). The key difference between RBAC and ACL is in how network or system resources are accessed. RBAC was a considerable improvement over ACLs as it is more scalable and flexible (Coyne and Weil, 2013). Although ACLs are still used, they have not led to established standards, whereas RBAC was standardized and is commonly used for identity and access management (IAM) in systems.

In the 2010 study, the benefits from RBAC adoption over 1992-2009 are estimated using rate of technology adoption and aggregating benefits across firms and industries. The comparison of costs and benefits gives net economic benefit from RBAC adoption. The efforts led by NIST resulted in time and cost savings for companies that adopted RBAC. Gallaher et al (2002) note that through efforts of NIST, adoption of RBAC standards was accelerated, thus accelerating technology innovation in a broad market.

In purely economic terms, NIST's expenditure over a decade (1992-2002) amounts to $2.6 million where it participated in development of the early RBAC model. Costs associated with deploying it over 2002-2006 was estimated to be $6.2 million and considers software developers' costs needed to incorporate RBAC functionality in systems of adopting companies. These costs were approximated using mean-wage rates and number of developer work-hours. The data was obtained from the US Bureau of Labor Statistics[2]. The developers' costs would have been $6.8 million without participation of NIST. Interestingly, actual costs were only 92% of what they would have been. This cost reduction combined with the early release of RBAC standards by a year resulted in a growth of 36 % in the number of companies that adopted RBAC over 1995-2009. The empirical analysis of RBAC adoption through I-MIAPS model is presented in Table 2.

| Components | Variables |
|---|---|
| **Model** | Decision to participate = $f$ (Costs (C), Influence ($I_t$), Market Innovation Advantage ($M_t$)) |
| **Cost (C)** | Initial investment by NIST in RBAC initiatives = $2.6 million<br><br>Software developer costs = $6.2 million<br><br>Software developer costs without NIST initiatives = $6.8 million |
| **Market Innovation Advantage (M)** | Net Economic Benefit = $0.6 million of $1.14 million in 2021$ |
| **Standards Life cycle (t)** | Reduction in RBAC adoption by 1 year<br><br>Steady growth rate from 1992-2002 because of standards development efforts by NIST. |

---

[2] https://www.bls.gov/ - U.S. Bureau of Labor Statistics

| | |
|---|---|
| **Influence ($I_t$)** | NIST's participation resulted in an increase in RBAC knowledge and early release of the standard by 1 year. This influenced firms seeking an alternative to ACL to rapidly adopt the RBAC standard. |
| | From 1995 when the RBAC penetration rate was just under a 4%, it grew to about 11% in 2002, 13% in 2004, and 41% in 2009. |

**Table 2:** *I-MIAPS Model for RBAC*

NIST's participation resulted in cost reduction while deploying RBAC standards and shortened time to achieve its broad adoption. The development of standards thus benefitted the industry, accelerated technology innovation, and increased efficiency and quality of products and services.

# E. Conclusion

In current global business environment, a significant number of companies participate in technical standardization. The explosive growth of ICT sector means that number of standardization efforts continues to expand. As a result, a decision support framework to model engagement in standardization is needed, to improve efficiency of standardization and accelerate its results. We build a decision model that considers tangible effects of standards on innovation in a technology area and resulting benefits for participating firms in terms of growth percentages and dollars. We incorporate effect of time and consider likelihood of success for a particular standard. In doing so, we also consider market innovation advantage and current and future influence a standard may exert in the technology space. Through our I-MIAPS model, we aim to understand framework, process, and components of standards participation. Using this understanding we provide a decision-making model that can drive a firm's participation in standard development. This model can help companies decide whether standards participation would be beneficial.

Future work will further develop this model by analyzing more complex use cases, e.g., 5G or 6G and other ICT related standards and use cases outside of the standardization space. We will test this model rigorously with standards currently in their advanced iteration. We will examine economic effects of international standardization, when, for example specifications developed by an industry consortium are taken to an international standards body (ISO), with future models examining different lifecycle trajectory of standards, including industry consortia, regional standards organizations, and international standard bodies. We will use a game theoretic approach to validate the model in future work. Using extensions of the simple decision model developed by us, we will analyze potential benefits of participation versus non-participation in standards for more complex cases. Similar approaches using Nash equilibrium were followed by Layne-Farrar et al. (2014) to investigate effect of multiple participating firms on investment decisions related to standards development. We will run additional simulations and case studies using data related to successful and discontinued standards. Using insights from these simulations, we aim to address more general issues in economics of standardization.

# F. References